\documentclass[twocolumn,showpacs,preprintnumbers,
               superscriptaddress]{revtex4}
\usepackage{graphicx,epsf}

\newcommand{\km}{(k-\mu)}

\newcommand{\beq}{\begin{eqnarray}}
\newcommand{\eeq}{\end{eqnarray}}

\newcommand{\btem}{\bibitem}

\begin{document}

\begin{flushright}
~\\
~\\
~\\
\vspace*{-2cm}
YITP-01-76, ~KUNS-1745
\end{flushright}
\vspace*{-.8cm}

\draft
\title{Precursor of  Color Superconductivity
 in Hot Quark Matter}

\author{M. Kitazawa}
\affiliation{Department of Physics,
 Kyoto University, Kyoto 606-8502, Japan}

\author{T. Koide}
\affiliation{Yukawa Institute for Theoretical Physics,
Kyoto University, Kyoto 606-8502, Japan}

\author{T. Kunihiro}
\affiliation{Yukawa Institute for Theoretical Physics,
Kyoto University, Kyoto 606-8502, Japan}

\author{Y. Nemoto\footnote{
Present address; RIKEN BNL Research Center, BNL, Upton, NY 11973.}}
\affiliation{Yukawa Institute for Theoretical Physics,
Kyoto University, Kyoto 606-8502, Japan}

\begin{abstract}
We investigate possible precursory phenomena of color superconductivity
in quark matter at finite temperature $T$ with use of a low-energy
 effective theory of QCD.
It is found that the fluctuating pair field exists with a prominent 
strength
even well above the critical temperature $T_c$. We show that the 
collective pair field has a complex energy located in the second
Riemann sheet, which approaches the origin as $T$ is lowered to $T_c$.
We discuss the possible relevance of the precursor to
the observables to be detected  in heavy  ion collisions.
\end{abstract}


\maketitle

It is an intriguing subject to explore how the QCD vacuum changes
 in hot and dense hadronic matter\cite{HK,BR}. 
The recent lattice QCD simulations
\cite{karsch} show that the QCD vacuum undergoes
 a phase transition to a chirally restored and deconfined phase
 at as low temperature ($T$) as $T\sim 170$ MeV.
Although it is still a great challenge to perform 
a Monte Carlo simulation with a finite chemical potential $\mu$ on
the lattice QCD with SU(3) color group, 
we believe that a chirally restored and 
deconfined system is realized in dense hadronic matter
with large $\mu$.

It is to be noticed that 
a system with a finite 
$\mu$ at low $T$ can have a Fermi surface and the system may be
described as a Fermi liquid composed of quarks.
In accordance with 
the validity of the diquark-quark  picture of 
baryons\cite{vogl},
the quark-quark interaction is attractive in some specific 
channels. Then  the existence of
 the Fermi surface gives rise to a Cooper 
instability with respect to the formation of the diquark or
Cooper pair in the most attractive channel, and the system is 
rearranged to a superconducting phase where
 the color-gauge symmetry is dynamically broken\cite{ref:Bar}.
Such a color superconductivity (CSC) has recently acquired a renewed
interest, because the resulting gap $\Delta$  was shown 
to  be as large as 100 MeV\cite{ref:NI,NI2}.
Furthermore the  color and flavor degrees of 
freedom of quarks give a fantastically rich structure in dense quark matter
with CSC\cite{review}:
Such a quark matter may be realized in the deep interior of 
neutron stars (NS), and  characteristic phenomena observed
for NS such as  the glitch phenomena
 might be attributed to the existence of 
the CSC in the NS.

Can CSC be relevant to experiments in the laboratory on the
earth?  The experiment using the available facilities,
such as RHIC using high energy heavy ion collisions seems unfortunately 
not fit for the problem, because the matter produced
by RHIC is almost baryon-free with
much higher $T$ than the critical temperature $T_c\sim 50$-60
 MeV of CSC. The purpose of the present paper is 
to explore possible precursory phenomena of CSC at  $T>T_c$, i.e., in the 
Wigner phase and show that the fluctuation of the  pair field is 
  significant at $T$ well above $T_c$, 
which thereby might affect observables to be detected by the
heavy ion collisions with  large baryon stopping\cite{ref:GSI}.

Since we are interested in relatively low $T$ and $\rho$ region,
we adopt a low-energy effective 
theory of QCD as was done in \cite{ref:NI,bergraj} where 
an instanton-induced interaction is employed.
We remark here that a simplified version of the instanton-induced 
interaction in the two-flavor case takes the form of 
 Nambu-Jona-Lasinio (NJL)  model\cite{njl}
which can be also derived by a Fierz transformation 
of the one-gluon exchange interaction
with heavy-gluon approximation; see
\cite{vogl,ref:K,HK,ebert}. Indeed it was shown later \cite{SKP} 
that the physical content given in \cite{bergraj} can be nicely
 reproduced by the  simple NJL model.
Therefore we  simply take
 the following NJL model as an effective theory, 
\begin{eqnarray}
\label{eqn:Lag}
{\cal L} &=& \bar{\psi}i/\hspace{-2mm}\partial \psi 
+ G_{S}[(\bar{\psi}\psi)^2
+ (\bar{\psi}i\gamma_{5}\vec{\tau} \psi)^2] \nonumber \\
&&+ G_{C}(\bar{\psi}i\gamma_{5}\tau_{2}
\lambda_{2}\psi^{C})(\bar{\psi}^{C}i\gamma_{5}\tau_{2}\lambda_{2}\psi),
\label{eqn:lagr}
\end{eqnarray}
where $\psi^{C} \equiv C\bar{\psi}^{T}$,
with $C = i\gamma^{2}\gamma^{0}$ being the charge conjugation 
operator.
Here we have confined ourselves in the 
two-flavor case in the chiral limit;
 $\tau_{2}$ and $\lambda_{2}$ are the second component of the 
Pauli and Gell-Mann matrices representing 
the flavor SU(2) and color SU(3), respectively.
The scalar coupling constant $G_{S}$ 
and the three dimensional momentum cutoff $\Lambda$ 
are determined so that the 
pion decay constant $f_{\pi}=93$ MeV and the 
chiral condensate $\langle \bar{\psi}\psi \rangle
=(-250 {\rm MeV})^{3}$ are reproduced in the chiral limit\cite{ref:K};
$G_{S} = 5.01{\rm GeV}^{-2}$ and $\Lambda = 650$ MeV.
As for the diquark coupling constant $G_{C}$,
we shall fix
$G_{C}$ at $3.11{\rm GeV}^{-2}$ \cite{SKP}  in this paper.
We have checked  that the qualitative features of the results 
do not change for  $G_{C}$ varied in the range used in the 
literature\cite{review,SKP}.

The phase diagram 
determined in the mean-field
approximation with (\ref{eqn:Lag}) is shown in Fig.1: 
The dashed line denotes the critical line
for a second order phase transition
and the solid line first order phase transition.
The critical temperature of the chiral transition at $\mu=0$
is $T=185$MeV, while the critical chemical potential
of chiral-CSC transition at $T=0$ is $\mu=316$MeV.
The tricritical point where the order of the chiral transition changes
from second to first order is located at about $(T,\mu)=(70,280)$MeV.
This phase diagram essentially 
coincides with those given in \cite{bergraj,SKP}.
\begin{figure}[tbp]
\begin{center}\leavevmode
\epsfxsize=6cm
\epsfbox{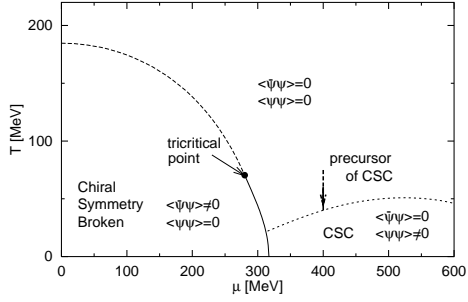} 
\caption{The calculated phase diagram in $T$-$\mu$ plane 
in our model.
The Solid and dashed line denote the critical line 
 of  a first and second order phase transition, 
respectively.
}
\end{center} 
\end{figure}

Given the phase diagram for the equilibrium state,
we  now proceed to examine fluctuations of the color Cooper pairs
in the normal phase at $T>T_c$.
We shall  obtain possible
 collective states and their strength functions in the linear
response theory. 
We suppose that the following  external pair field is applied to
 the system,
$
\Delta^*_{ex}(x) = \Delta^*_{ex} e^{i {\bf k \cdot x}}e^{-i\omega_n \tau},
$
 which is tantamount to adding the external Hamiltonian
$H_{ex} = \int d{\bf x} \Delta^*_{ex}(x)
(\bar{\psi}^C i \gamma_5 \tau_2 \lambda_2 \psi ) + \mbox{h.c.}$.
Then a color Cooper-pair field is dynamically induced, as  given by
$\Delta^{*}_{ind}({\bf x},\tau)
= -2G_{C}\langle \bar{\psi}({\bf x},\tau)\tau_{2}\lambda_{2}i\gamma_{5}
\psi^{C} ({\bf x},\tau) \rangle$, 
where the bracket denotes a statistical average. 
Using  the anomalous Green function defined by
$F^{\dagger}({\bf x},\tau;{\bf x}',\tau')
= -\langle T_{\tau}\{ \psi^{C}({\bf x},\tau)\bar{\psi}({\bf x}',\tau') \}
\rangle$,
 the induced pair field is expressed as 
\beq
\label{eqn:scc}
\Delta^{*}_{ind}({\bf x},\tau)=
-2G_{C}{\rm Tr}~\tau_{2}\lambda_{2}i\gamma_{5} F^{\dagger}({\bf x},
\tau;{\bf x},\tau_{+}),
\eeq
where
${\rm Tr}$ denotes
the trace taken over the color, flavor and spinor indices.
$ F^{\dagger}({\bf x},\tau;{\bf x}',\tau')$
 satisfies the following differential equation 
\begin{eqnarray}
\lefteqn{[-\gamma^{0}\frac{\partial}{\partial \tau} - K_{0}
-\mu\gamma^{0}] F^{\dagger}({\bf x},\tau;{\bf x}',\tau')}\nonumber \\
&&\hspace*{1cm}= \Delta^{*}_{tot}({\bf x},\tau)
\tau_{2}\lambda_{2}i\gamma_{5}G({\bf x},\tau;{\bf x}',\tau'),\label{eqn:AGEQ}
\end{eqnarray}
where $K_{0} = -i\vec{\gamma}\cdot\vec{\nabla} + M$ and 
$M = -2G_{S}\langle \bar{\psi}({\bf x},\tau)\psi({\bf x},\tau)\rangle$
is the dynamical mass. Here,
$\Delta_{tot}^{*}({\bf x},\tau)
=\Delta^{*}_{ex}({\bf x},\tau)+\Delta^{*}_{ind}({\bf x},\tau)$
is the total pair field to induce the pair field:
Notice the self-consistent nature of the problem.
 $G({\bf x},\tau;{\bf x}',\tau')$ is 
the usual non-anomalous Matsubara Green function.
To see a linear response of the system, one may  replace 
$G({\bf x},\tau;{\bf x}',\tau')$  in 
Eq.(\ref{eqn:AGEQ})
with a free Matsubara Green function
$G^{(0)}({\bf x},\tau;{\bf x}',\tau')$; thus we have
$ F^{\dagger}({\bf x},\tau;{\bf x}',\tau_{+})=$
$(-\gamma^{0}\frac{\partial}{\partial\tau}-K_{0} -\tilde{\mu}\gamma^{0})^{-1} 
\times \Delta^{*}_{tot}({\bf x},\tau)
\tau_{2}\lambda_{2}i\gamma_{5}G^{(0)}({\bf x},\tau;{\bf x},\tau_{+})$.
Inserting this expression into Eq.(\ref{eqn:scc}) and
 performing the Fourier transformation, we have
\begin{eqnarray}
\Delta^{*}_{ind}({\bf k},\omega_{n})
\equiv -G_{C}\Delta^{*}_{tot}({\bf k},\omega_{n}){\cal Q}({\bf
k},\omega_{n}),
\label{eqn:fou}
\end{eqnarray}
where
\begin{widetext}
\begin{eqnarray}
{\cal Q}({\bf k},\omega_{n}) 
&=& -2N_{f}(N_{c}-1)
\int \frac{d^3 {\bf k}'}{(2\pi)^3}\frac{1}{E_{\bf k'}E_{\bf k-k'}}
\nonumber \\
&& \times[ 
2(f^{-}(E_{\bf k'})(1-f^{+}(E_{\bf k-k'}))
-f^{+}(E_{\bf k-k'})(1-f^{-}(E_{\bf k'})))
\frac{-E_{\bf k'}E_{\bf k-k'}-{\bf k}'\cdot({\bf k-k'})+M^2}
{-i\omega_{n}-2\mu-E_{\bf k-k'}+E_{\bf k'}}\nonumber \\
&& -(1-f^{-}(E_{\bf k'})-f^{-}(E_{\bf k-k'}))
\frac{-E_{\bf k'}E_{\bf k-k'}+{\bf k}'\cdot({\bf k-k'})-M^2}
{-i\omega_{n}-2\mu+E_{\bf k-k'}+E_{\bf k'}} \nonumber \\
&&+(1-f^{+}(E_{\bf k'})-f^{+}(E_{\bf k-k'}))
\frac{-E_{\bf k'}E_{\bf k-k'}+{\bf k}'\cdot({\bf k-k'})-M^2}
{-i\omega_{n}-2\mu-E_{\bf k-k'}-E_{\bf k'}}
].\label{eqn:Q}
\end{eqnarray}
\end{widetext}
Here, $E_{\bf k} = \sqrt{{\bf k}^2 + M^2}$ and $f^{\pm}(E) =[\exp(\beta
(E \pm \mu))+1]^{-1}$ with $\beta=T^{-1}$.
The first term in the brace on the r.h.s. of Eq.(\ref{eqn:Q}) is 
originated from the process
$\Delta^{*}_{tot} + q \rightarrow \bar{q}$ and 
its inverse process $\bar{q} \rightarrow \Delta^{*}_{tot} + q$.
The second and third terms can be interpreted similarly in terms
of kinetic processes.
Inserting $\Delta_{tot}^{*}({\bf k}, \omega_n)
=\Delta^{*}_{ex}({\bf k},\omega_n)+\Delta^{*}_{ind}({\bf k},\omega_n)$
into Eq.(\ref{eqn:fou}),
we obtain the induced pair field in the linear response
 to  the external field in the imaginary time as
\begin{eqnarray}
\Delta^{*}_{ind}({\bf k},\omega_{n})
=
\frac{-G_{C}{\cal Q}({\bf k},\omega_{n})}{1+G_{C}{\cal Q}({\bf k},\omega_{n})}
\Delta^{*}_{ex} 
\equiv {\cal D}({\bf k},i\omega_{n})\Delta^{*}_{ex}. \label{eqn:res}
\end{eqnarray}
This is a central result in the present work,
on which the rest of discussions is based.
The proportionality constant in 
Eq.(\ref{eqn:res}) converted in the real-time domain
is called the  response function or retarded
 Green's function $D^{R}({\bf k},\omega)$, which  is obtained by 
the analytic continuation of $ {\cal D}({\bf k},\omega)$ to the upper half
of the complex energy plane as
$D^{R}({\bf k},\omega) = {\cal D}({\bf k},\omega + i\eta).$

Setting  $M=0$, let us examine precursory 
collective modes above the CSC
 phase when approaching the critical point
 as shown by the arrow in Fig. 1.
Then we notice the following equality holds 
\begin{eqnarray}
\label{eqn:critical}
\left.1+G_{C}{\cal Q}({\bf 0},0)\right|_{T=T_{c}}=0,
\end{eqnarray}
owing to 
the self-consistency condition for the diquark condensate 
at $T=T_c$. This equality is the origin to lead to 
the various precritical phenomena for CSC
, as we will see shortly.
Here it should be stressed that although the following discussion
 is seemingly relied heavily on the fact that phase transition is second
order, 
the nature of the collective mode to be discussed below 
will not be altered even when the
phase transition is weak first order.

The spectral function which is the measure of the excitation 
strength of the fluctuating pair field  is given by
\begin{equation}
\rho({\bf k},\omega)= -\frac{1}{\pi}{\rm Im}D^{R}({\bf k},\omega). 
\end{equation}
The temperature dependence of the spectral function 
at $\mu=400$ MeV and zero momentum transfer (${\bf k}=0$) 
is given in Fig.2(a). One can see a prominent peak
moving toward the origin
 grows as the temperature is lowered toward $T_c$ as indicated by 
 the arrow in Fig.1.
In the experimental point of view,
 it is interesting that the peak survives at $T$ even well 
above $T_c$ with $\epsilon\equiv (T-T_c)/T_c\sim 0.2$.
This means that the precritical region of CSC is one to two 
order larger in the unit of $T_c$ than that in 
electric superconductors.
Of course, this is related to the difference of the ratio
$\omega_D/E_F$ and $\Lambda/E_F$ where $\omega_D$ is the Debye frequency
and $E_F$ is the Fermi energy; NJL model is similar to the attractive
Hubbard model than to BCS model\cite{loktev}.
The result  for  $\mu=500$ MeV  is 
shown in Fig.2(b);  one sees that although 
the growth of the peak becomes relatively moderate, 
the qualitative features do not change.
\begin{figure}[tbh]
\begin{center}\leavevmode
\epsfxsize=8cm
\epsfbox{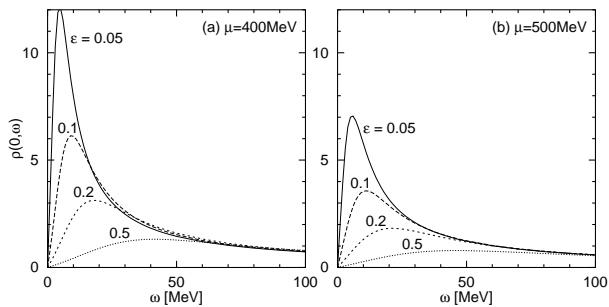} 
\caption{The spectral function  for the pair field at $T>T_c$ 
with $\epsilon \equiv (T-T_c)/T_c = 0.05$,$0.1$,$0.2$ and $0.5$ 
at $\mu = 400$MeV (a) and $\mu = 500$MeV (b).
}
\end{center} 
\end{figure}

The existence of the peak with the narrow width suggests 
an existence of a well-defined collective mode as an
elementary excitation in the Wigner phase.
The existence  and the dispersion relation 
$\omega=\omega(k)$ of a collective excitation in a isotropic
system  is examined by 
searching  possible poles of 
the response function, i.e., solving the equation
$1+G_{C}{Q}({k}, \omega) = 0$. Here
 ${Q}({k}, \omega)$ is the analytic continuation of
${\cal Q}({k},\omega_{n})$ to the upper half plane.
The existence of a pole
$\omega=\omega(k)$ for given ${k}$ means that 
the pair field $\Delta_{ind}({k},\omega(k))$ can be created
even with an infinitesimal external field $\Delta^{*}_{ex}$
as seen from Eq.(\ref{eqn:res}),
 which physically means that the system admits  spontaneous
excitation  of a collective mode 
with the dispersion relation $\omega=\omega(k)$.
Since the energy of the collective mode is
found to be the lowest at ${k}=0$, let us 
obtain $\omega(0)\equiv z$.
One should notice that 
the gap equation  Eq.(\ref{eqn:critical})  
implies that $z=0$ is a solution at $T=T_c$, i.e., 
there exists a zero mode at the critical point.
The singular behavior seen in the spectral function
near the critical point in Fig.2 is caused by the presence of
 the zero mode.
Conversely, one may  take 
the emergence of the zero mode as the critical condition
 for the color superconductivity. This condition is actually long 
known in the condensed matter physics and  called Thouless 
criterion\cite{ref:Thoul}.

Notice that the pole should be located in the lower half 
plane otherwise the system is unstable for the creation of the
collective mode. Because of a cut along the real $z$ axis, 
the analytic continuation of $Q(0, z)$ to 
the lower half plane leads  to the second Riemann sheet; 
\beq
Q(0, z)&=&
\frac{2N_f(N_c-1)}{\pi^2}\big[
{\rm P}\int_{-\Lambda}^{\Lambda}
 k^2 dk\frac{\tanh \frac{\km}{2T}}{z-2\km}\nonumber \\
 & &-\frac{1}{2}2\pi i(\mu+\frac z2)^2\tanh \frac{z}{4T}
\big],
\eeq
where P denotes the principal value and
 the term in the second line is the additional term
for $z$ to be in the second sheet.

A numerical calculation shows that 
there indeed exists a pole in the lower half plane
 for $T>T_c$.
Fig.3 shows how the pole moves as $T$ is lowered toward $T_c$ for
$\mu=400$ and $500$ MeV.
One sees that the pole approaches the origin 
as $T$ is lowered toward $T_c$.
Such a mode whose energy tends to vanish as the system approaches the 
critical point of a phase transition is called a soft mode\cite{HK}.
Thus
we have found the pole for the soft mode for CSC in hot quark matter.
As far as we know, this is the first time
when the identification has been done of the softening 
pole in the complex energy plane 
for the superconductivity  including the condensed
matter physics\cite{ref:Thoul}. 
The significance of the existence of  such a soft mode  at $k=0$
lies in the fact that it causes a long-range correlations  
for the  color Cooper pairs  in the Wigner phase,
 which will leads to a 
singular behavior of various observables such as the transport
coefficients, as will be discussed later.
We remark  that
 such a soft mode in hot and dense quark matter
was first discussed in \cite{ref:HK} in relation to the chiral
transition.

Another characteristic feature of the soft mode is that 
the absolute value of the imaginary part $\omega_i$ of the pole 
is larger than that of the real part $\omega_r$ for both $\mu$.
This feature may have an important implication for an effective
equation describing the dynamical phase transition for CSC:
The dynamical behavior of the order parameter near $T_c$
is  well described by a non-linear diffusion equation which is known as
the time-dependent Ginzburg-Landau (TDGL) equation  \cite{ref:Cyr}.
We however remark here that by setting $k=0$,
 the so called Landau damping due to the particle-hole excitation 
is not incorporated in the present calculation.
Fig.3  shows that  the ratio $\omega_i/\omega_r$ becomes 
larger for the larger $\mu$.
This is plausible because the larger the Fermi energy, the larger the 
density of states near the Fermi surface, which
implies that the volume of 
the phase space for the decay of the pair field increases
 with $\mu$ at low $T$.

\begin{figure}[tbh]
\begin{center}\leavevmode
\epsfxsize=6cm
\epsfbox{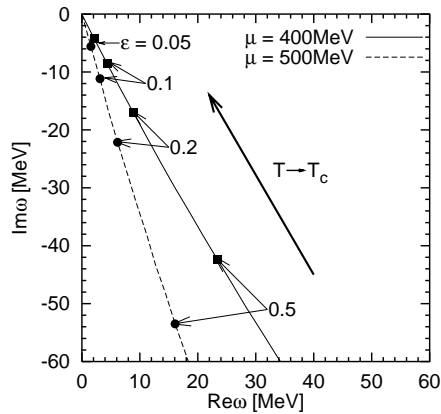} 
\caption{The pole of the precursory soft mode 
for $\mu = 400$ MeV and $500$MeV.
The  numbers attached to dots denote $\epsilon=(T-T_c)/T_c$.
}
\end{center} 
\end{figure}

It is interesting that  the so called pseudogap in
 the density of states observed in  high-$T_c$ superconductors
 (HTSC) as a characteristic precursory 
phenomenon\cite{tim} might be attributed to 
fluctuations of the Cooper pairs\cite{levin}.
Thus 
the present results strongly
suggests such a pseudogap phase exists also 
for CSC  in a rather broad region of $T$ above $T_c$.
To establish the pseudogap phase,
one must calculate the density of states from the single particle
Green's function\cite{loktev,levin}, which we hope to  
report in a future publication.
Experimentally, a measurement of the single-particle quark propagator
 is useful for obtaining the information on the density of states for 
the single quark states, as was done for HTSC by 
a kind of the quasi-elastic scattering by photon,
 called the angle-resolved 
photo-emission spectroscopy (ARPES)\cite{tim}.

The fluctuation of the 
pair field in  metals above $T_c$ affects several transport
coefficients
like the electric conductivity (EC), the phonon absorption 
coefficient and the specific heat\cite{ref:AL}.
An anomalous excess of the 
electric conductivity $T_c$ is known as 
paraconductivity (PC)\cite{ref:AL}. 
The direct analogy thus leads to an expectation of the enhancement
of color conductivity above $T_c$ in our case, which we call
color-paraconductivity (CPC):
In fact, the PC for the metal superconductivity comes from the fact
that the effective dynamics is given by  TDGL equation giving rise to a
zero mode at momentum $k=0$ at $T=T_c$\cite{ref:AL},
 which is just the behavior extracted from the behavior of 
our soft mode as shown in Fig.3. 

More interesting  is that  the electric current 
can also have an anomalous enhancement in the Wigner phase due  to 
the color pair-field fluctuations:
 This is because
the propagator of  photons (and also gluons)  at $T>T_c$
is modified by the fluctuating pair-field
 as in the usual superconductors\cite{ref:AL}.
This implies that photons and hence dileptons from the system
 can carry some information of the fluctuation of the
color pair field.

In short,
we have explored  precursory phenomena
for the  color superconductivity (CSC)
in quark matter at $T\not=0$; 
the fluctuating pair field exists with a prominent 
strength  even well above $T_c$.
We have also discussed 
the possible observability of CSC through the fluctuations 
by  heavy  ion collisions with large baryon stopping 
using an analogy with the precritical phenomena 
known  in the metal and the high-$T_c$ superconductors.

We thank M. Alford for his interest in our work and
suggesting the possible relevance of 
our work to \cite{bord}.
O. Tchernyshyov is gratefully acknowledged 
for his encouragement and  useful comments 
on the  relevance of the pseudogap
phenomena in condensed matter physics\cite{Tch} to our case.
We are indebted to 
 M. Asakawa for indicating a misleading statement present in the 
original manuscript.
T.Kunihiro is partially supported by the Grants-in-Aid of 
the Japanese Ministry of Education, Science and Culture 
(No. 12640263 and 12640296).

\end{document}